\documentstyle[prl,twocolumn,aps]{revtex}
\RequirePackage{epsfig}
\def\Journal#1#2#3#4{{#1} {\bf #2}, #3 (#4)}
\def\PR{Phys. Rev.}
\def\PRL{Phys. Rev. Lett.}
\def\PRA{Phys. Rev. A}
\def\JMP{J. Math. Phys.}
\def\CPL{Chem. Phys. Lett.}
\def\IJQCQCS{Intern. J. Quantum Chem. Quantum Chem. Symp.}

\def\SC{Science}
\def\JPB{J. Phys. B: At. Mol. Opt. Phys.}
\def\EPJD{{\em Eur. Phys. J.} D}
\def\Nature{\em Nature}
\begin{document}
\draft
\title {Measurement of one-particle correlations and momentum
distributions\\ for trapped 1D gases}
\author{M. D. Girardeau\thanks{Email: girardeau@optics.arizona.edu}
and E. M. Wright\thanks{Email: Ewan.Wright@optics.arizona.edu}}
\address{Optical Sciences Center and Department of Physics,
University of Arizona, Tucson, AZ 85721}
\date{\today}
\maketitle
\begin{abstract}
van Hove's theory of scattering of probe particles by a
macroscopic target is generalized so as to relate the differential
cross section for atomic ejection via stimulated Raman transitions
to one-particle momentum-time correlations and momentum
distributions of 1D trapped gases. This method is well suited to
probing the longitudinal momentum distributions of 1D gases in
situ, and examples are given for bosonic and fermionic atoms.
\end{abstract}
If an ultracold atomic gas is probed by high-energy particle
scattering so that the struck boson recoils with an energy large
compared with typical interactions in the target system then
measurement of the differential scattering cross section leads to
determination of the dynamic structure factor \cite{HP,SP,ZPSS}.
There are some situations not amenable to such a treatment. For
example, it follows from the Fermi-Bose mapping theorem \cite{map}
that the dynamic structure factor of a one-dimensional (1D) system
of hard core bosons (Tonks gas \cite{Ols98,PetShlWal00}) in a
tightly confined atom waveguide
\cite{HinBosHug98,ThyWesPre99,BonBurDet00} is identical with that
of the corresponding ideal Fermi gas, although their momentum
distributions are quite different. Here we introduce another
approach based on a generalization \cite{Micha,GI} of the classic
van Hove theory \cite{vanHove} of scattering of probe particles by
a macroscopic target. We apply this very general approach to
stimulated Raman transitions between two different hyperfine
levels of a 1D trapped gas, for example, a Bose-Einstein
condensate (BEC). Raman transitions have been demonstrated as a
mechanism for outcoupling of coherent atom beams from a BEC
\cite{MHS,Hagley-etal,Edwards-etal}, and Bloch {\it et al.}
\cite{BloHanEss00} have employed radio-wave output coupling to
measure the spatial coherence of trapped atomic gases.  Here we
propose using Raman outcoupling to determine single-atom
momentum-time correlations and momentum distributions of 1D
trapped gases. The relevant reactive scattering process in the
case of a sodium BEC experiment is \cite{Hagley-etal}
$photon+BEC\rightarrow photon^{'}+boson^{'}+BEC^{'}$ where
$photon$ denotes a photon absorbed from an incident laser beam,
$photon^{'}$ denotes a photon emitted into a second overlapping
laser beam via a stimulated Raman transition, $boson^{'}$ denotes
an ejected atom in the untrapped $m=0$ magnetic sublevel, which
originally belonged to an $N$-atom BEC of atoms in the $m=-1$
sublevel, and $BEC{'}$ denotes an $(N-1)$-atom excited state of
the residual BEC.  The same scheme applies to the experiments on
Bragg spectroscopy of sodium BECs \cite{SteInoChi99,StaChiGor99}
but there the internal state of the scattered atom $boson^{'}$
remains in the trapped state $m=-1$.  It is this distinction that
allows our scheme to yield the momentum distribution directly by
examination of the ejected atoms in contrast to Bragg spectroscopy
which yields the dynamic structure factor.

Another recent proposal \cite{KS} for measuring the one-particle
density matrix via the reactive scattering process
$impurity+BEC\rightarrow impurity+boson+BEC^{'}$ simultaneously
assumes a {\em high} energy incident impurity (Born approximation)
and an S-wave pseudopotential impurity-boson interaction justified
only for {\em low} energy collisions, and requires measurement of
a differential cross section involving {\em two} outgoing
particles. Some recent work \cite{GM} on the use of
photoionization measurements for determination of correlation
functions of the Bose field used a different scheme (tightly
focused laser beam, measurement of photoelectron energy but not
angular distribution) with emphasis on temporal rather then
spatial correlations.

\begin{figure}
\epsfxsize 2.2in \epsfbox{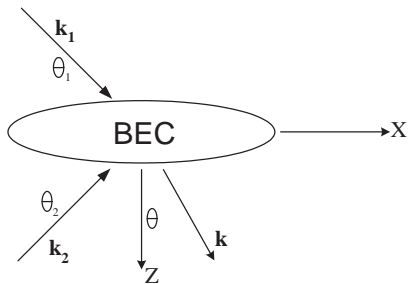} \caption{Geometry
of the Raman outcoupling} \label{Fig:one}
\end{figure}

{\it Derivation}: A crucial point \cite{GI} is that in order to
probe {\em one}-particle correlations one should use a {\em
reactive} scattering process which effectively {\em removes} one
particle from the target system; this leads to the {\em hole
propagator} and a related one-particle momentum-time correlation
function. For the case of the above-described Raman process, we
use the following simple choice of asymptotic initial and final
states: $|i\rangle=|\mbox{\ laser}\rangle
|\mbox{B}_{i}\rangle=\hat{U}|\mbox{B}_{i}\rangle$ and
$|f\rangle=\hat{a}_{{\bf k},m=0}^{\dagger}|\mbox{\ laser}\rangle
|\mbox{B}_{f}\rangle=\hat{a}_{{\bf k},m=0}^{\dagger}\hat{U}
|\mbox{B}_{f}\rangle$. The laser state vector is a single-mode
coherent state with respect to each of the two Raman beams, one
with wave vector ${\bf k}_{1}$ and polarization index
$\lambda_{1}$ and the other with wave vector ${\bf k}_{2}$ and
polarization index $\lambda_{2}$. Its most convenient
representation for our purposes is $|\mbox{\
laser}\rangle=\hat{U}|0\rangle$ where the unitary operator
$\hat{U}=e^{\hat{F}}$ with $\hat{F}=c_{1}\hat{b}_{{\bf
k}_{1}\lambda_{1}}^{\dagger} +c_{2}\hat{b}_{{\bf
k}_{2}\lambda_{2}}^{\dagger}-\mbox{h.c.}$. The mean photon
occupation numbers are $n_{j}=|c_{j}|^2$ and the occupation number
dispersions are $\sqrt{n_{j}}$, negligible compared to $n{_j}\gg
1$. For concreteness in notation we consider the case of a sodium
BEC target but stress that the treatment is not restricted to
bosonic atoms or equilibrium states. Then $|\mbox{B}_{i}\rangle$
is the $N$-boson state vector describing $N$ magnetically trapped
bosons in their electronic ground state and magnetic sublevel
$m=-1$, including all effects of their mutual interactions,
$\hat{a}_{{\bf k},m=0}^{\dagger}$ creates an atom with wave vector
${\bf k}$ in the untrapped $m=0$ magnetic sublevel, and
$|\mbox{B}_{f}\rangle$ is an $(N-1)$-boson final state of the
residual BEC after the $m=0$ atom has been ejected. The geometry
of this process is shown in Figure \ref{Fig:one} where we show a
cigar shaped BEC of length $L$ along the longitudinal direction
$x$. We assume that the BEC is tightly confined in the transverse
plane ${\bf r}_T=(y,z)$ using an atom waveguide. At resonance the
energy difference $\hbar c(k_{1}-k_{2})$ between the photon
absorbed by the atom (number $1$) and that emitted (number $2$) is
equal to the energy splitting $\Delta$ between a trapped atom with
$m=-1$ and that of the untrapped atom with $m=0$ ($\Delta$
includes both the Zeemen splitting and difference in confinement
energy between the two states). The net momentum $\hbar({\bf
k}_{1}-{\bf k}_{2})$ transferred to the atom will be in the
forward direction $z$ as shown provided that the indicated angles
of the two beams satisfy
$\sin\theta_{2}/\sin\theta_{1}=k_{1}/k_{2}\approx 1$. The
requirement is then $\theta_{2}=\theta_{1}+\epsilon$ with
$\epsilon\approx[(k_{1}-k_{2})/k_{2}]\tan\theta_{1}\ll 1$. In the
design of experiments the polarizations $\lambda_1$ and
$\lambda_2$ of the two laser beams must be chosen appropriately,
and a suitable detuning must be used to inhibit spontaneous
transitions \cite{Hagley-etal}. Inclusion of these details changes
the differential cross section only by a multiplicative factor and
will not be considered here.

The appropriate Hamiltonian is
$\hat{H}=\hat{H}_{B}+\hat{H}_{0ph}+\hat{H}_{0a}+\hat{V}$
where $\hat{H}_{B}$ is the second quantized many-body Hamiltonian of
the trapped $m=-1$ BEC atoms including their mutual interactions
and $\hat{V}$ is the interaction of the $N$-atom system with the quantized
electromagnetic field. The asymptotic (free) Hamiltonians for the incident
photons and outgoing $m=0$ atoms are
$\hat{H}_{0ph}=\hbar c(k_{1}\hat{N}_{1}+k_{2}\hat{N}_{2})$ and
$\hat{H}_{0a}=\sum_{{\bf k}}[(\hbar^{2}k^{2}/2m)+\Delta]\hat{N}_{{\bf k},m=0}$
where $\hat{N}_{j}=\hat{b}_{{\bf k}_{j}\lambda_{j}}^{\dagger}
\hat{b}_{{\bf k}_{j}\lambda_{j}}$ and
$\hat{N}_{{\bf k},m=0}=\hat{a}_{{\bf k},m=0}^{\dagger}\hat{a}_{{\bf k},m=0}$
are the occupation number operators for photons and $m=0$ atoms in their
electronic ground state and $\Delta$ is the energy difference
between $m=-1$ and $m=0$ atoms. The coherent laser
state $|\mbox{\ laser}\rangle$ is not an exact eigenstate of $\hat{H}_{0ph}$,
but its energy dispersion is negligible since the photon numbers are large.

The derivation closely parallels our previous work \cite{GI}.
The exact transition rate from a specified initial state to a specified final
state is $(2\pi/\hbar)|\langle f|\hat{T}|i\rangle|^{2}\delta(E_{f}-E_{i})$
where $E_{f}-E_{i}=E_{\mbox{B}_{f}}-E_{\mbox{B}_{i}}-\hbar\omega$ with
$\hbar\omega=\hbar c(k_{1}-k_{2})-(\hbar^{2}k^{2}/2m)-\Delta$,
the energy transferred to the BEC; the momentum transferred is
$\hbar{\bf q}=\hbar({\bf k}_{1}-{\bf k}_{2}-{\bf k})$. Here
$\hat{T}$ is the exact transition operator
$\hat{T}=\hat{V}\hat{\Omega}^{+}=\hat{\Omega}^{-}\hat{V}$ where
$\hat{\Omega}^{\pm}$ are the M\"{o}ller wave operators $\hat{U}(0,\mp\infty)$
and $\hat{U}$ is the adiabatically switched evolution operator
\cite{Joachain}. The angle-energy doubly differential cross section is
then \cite{GI}
\begin{equation}
\frac{d^{2}\sigma}{d\Omega d\omega}=\rho(\theta,\phi)
\int_{-\infty}^{\infty}dt\langle{\hat{U}}^{-1}
\hat{T}^{\dagger}(t)\hat{N}_{{\bf k},m=0}\hat{T}(0)\hat{U}\rangle_{B}^{'}
\end{equation}
where $\langle\cdots\rangle_{B}$ denotes an average over a
statistical ensemble of BEC initial states $|\mbox{B}_{i}\rangle$
(reducing to an $N$-boson ground state average for temperature
$T\rightarrow 0$) and the prime indicates that this expression is
to be evaluated for an outgoing $m=0$ atom with ${\bf k}$ vector
in the direction $(\theta,\phi)$ and with magnitude
$k=\sqrt{(2m/\hbar^{2})[\hbar c(k_{1}-k_{2})-\Delta-\hbar\omega]}$
consistent with a specified energy transfer $\hbar\omega$ to the
BEC. Here $\rho(\theta,\phi)$ is the density of final states for
an observation distance $D$ from the 1D gas which is much larger
than the length $L$ of the trapped gas, $\phi$ is the angle of
atomic ejection in the transverse plane, and $\theta$ the angle of
ejection measured out of the transverse plane (see Figure
\ref{Fig:one}). $\hat{T}^{\dagger}(t)$ is the Heisenberg operator
\begin{equation}
\hat{T}^{\dagger}(t)=e^{it(\hat{H}_{B}+\hat{H}_{0ph}+\hat{H}_{0a})/\hbar}
\hat{T}^{\dagger}e^{-it(\hat{H}_{B}+\hat{H}_{0ph}+\hat{H}_{0a})/\hbar}
\end{equation}
and the BEC final states have been eliminated by an exact closure
summation
$\sum_{\mbox{B}_{f}}|\mbox{B}_{f}\rangle\langle\mbox{B}_{f}|=\hat{1}$
after introducing the integral representation of the energy delta
function \cite{Micha,GI,vanHove}.

The terms in $\hat{T}(0)=\hat{T}$ contributing
to the Raman process have the structure
\begin{equation}
\sum_{{\bf k}_{1}^{'}{\bf k}_{2}^{'}}
\hat{b}_{{\bf k}_{2}\lambda_{2}}^{\dagger}
\hat{a}_{{\bf k}_{2}^{'},m=0}^{\dagger}
({\bf k}_{2}\lambda_{2},{\bf k}_{2}^{'}|T|{\bf k}_{1}\lambda_{1},
{\bf k}_{1}^{'})\hat{a}_{{\bf k}_{1}^{'},m=-1}\hat{b}_{{\bf k}_{1}\lambda_{1}}
\end{equation}
To obtain the relevant terms in $\hat{T}^{\dagger}(t)$ we take
the hermitian conjugate of (3) and propagate each of its annihilation and
creation operators to time $t$ in accordance with (2). Heisenberg propagation
of $\hat{b}_{{\bf k}\lambda}^{\dagger}$ and $\hat{a}_{{\bf k}^{'},m=0}$
gives only trivial phases since the target BEC contains only atoms with
$m=-1$, whereas the propagation of $\hat{a}_{{\bf k}^{'},m=-1}$
is nontrivial and leads to inclusion of effects of energy and momentum
transfer to the target BEC. Noting that the T-matrix elements
$({\bf k}_{2}\lambda_{2},{\bf k}_{2}^{'}|T|{\bf k}_{1}\lambda_{1},
{\bf k}_{1}^{'})$ contain a momentum conservation factor
$\delta_{{\bf k}_{1}+{\bf k}_{1}^{'},{\bf k}_{2}+{\bf k}_{2}^{'}}$
and performing Wick's theorem algebra after evaluating the coherent laser
unitary transformation $\hat{U}^{-1}\cdots\hat{U}$ in Eq. (1), one finds
\begin{eqnarray}
\frac{d^{2}\sigma}{d\Omega d\omega} & = &
n_{1}(n_{2}+1)\rho(\theta,\phi) \nonumber\\&\times & |({\bf
k}_{2}\lambda_{2},{\bf k}|T|{\bf k}_{1}\lambda_{1}, {\bf k}-{\bf
k}_{1}+{\bf k}_{2})|^{2} \nonumber\\ & \times &
\int_{-\infty}^{\infty} \langle\hat{a}_{{\bf k}-{\bf k}_{1}+{\bf
k}_{2}}^{\dagger}(t) \hat{a}_{{\bf k}-{\bf k}_{1}+{\bf
k}_{2}}(0)\rangle_{B}^{'}e^{i\omega t}dt \label{exact}
\end{eqnarray}
where the annihilation and creation operators refer to atoms in
the BEC target, their $m=-1$ subscript having been dropped for
simplicity in notation. Physically, Eq. (\ref{exact}) reflects
that fact that outgoing $m=0$ atoms with momentum $\hbar{\bf k}$
originate from transfer of momentum $\hbar({\bf k}_{1}-{\bf
k}_{2})$ from the Raman photons to the target bosons of momentum
$\hbar({\bf k}-{\bf k}_{1}+{\bf k}_{2})$ and the differential
cross section at the corresponding angle is related to the number
of such target bosons.

{\em 1D trapped gases}: The expression (\ref{exact}) is exact. We
now introduce approximations appropriate to tightly confined 1D
gases where only the ground transverse mode is occupied
\cite{PetShlWal00}. When an atom is transferred by the Raman
transition from the $m=-1$ trapped state to the $m=0$ untrapped
state it is ejected and undergoes radial expansion in the
transverse plane ${\bf r}_T=(y,z)$ in addition to the momentum
kick $\hbar({\bf k}_1-{\bf k}_2)$ in the forward direction $z$
from the Raman process. In the limit that the transverse mode
$u_g({\bf r}_T)$ of the trapped atoms has a width $w_T$ much less
than both the length $L$ of the sample and the Raman laser
wavelength, the ejected atom wave function will predominantly
propagate out radially as a cylindrical wave concentrated along
the direction $\theta=0$.  The density of final states
$\rho(\theta,\phi)$ may then be replaced by a constant (we assume
the atoms can be detected before gravity significantly changes the
final density of states). In addition, writing the field operator
for the trapped target atoms as $\hat\psi({\bf r},t)=u_g({\bf
r}_T)\hat v(x,t)$ to reflect their single-transverse mode
structure, and setting ${\bf k}={\bf k}_T+{\bf x}k_x$ with
$k_x=k\sin(\theta)$ the longitudinal component of the wavevector,
we find, for example
\begin{eqnarray}
\hat{a}_{ {\bf k}-{\bf k}_{1}+{\bf k}_{2} }(t) &=& \int d^3{\bf r}
\hat\psi({\bf r},t)e^{ -i({\bf k}-{\bf k}_{1}+{\bf k}_{2})\cdot
{\bf r} }  , \nonumber \\ &=& \tilde{u}_g( {\bf
K}_T)\hat{c}(k_x,t) .
\end{eqnarray}
Here $\tilde u_g({\bf K}_T)$ is the Fourier transform over the
transverse plane of the transverse mode evaluated at ${\bf
K}_T=({\bf k}_T-{\bf k}_{1}+{\bf k}_{2})$, that may be
approximated by its value $\tilde u_g(0)$ at zero momentum under
tight confinement conditions $2\pi/w_T >> |{\bf k}_T-{\bf
k}_{1}+{\bf k}_{2}|$, and $\hat{c}(k_x,t)=\int dx\hat
v(x,t)\exp(-ik_x x)$ is the Fourier transform of the longitudinal
field operator. Putting this together in Eq. (\ref{exact}) and
treating the T-matrix elements as constants (see below) yields an
expression proportional to the Fourier transform of the
longitudinal one-particle momentum-time correlation function:
\begin{equation}
\frac{d^{2}\sigma}{d\Omega d\omega}\approx \alpha
\int_{-\infty}^{\infty} \langle\hat{c}^{\dagger}(k_x,t)
\hat{c}(k_x,0)\rangle_{B}^{'}e^{i\omega t}dt \label{FT}
\end{equation}
where $\alpha$ is a constant.  Thus, measurement of the doubly
differential cross section for the ejected atoms gives access to a
one-particle momentum-time correlation function of the trapped
atoms.

Further simplification is possible if measuring the energy
transferred to the BEC is not feasible, and only the angular
distribution is measured. We consider that the 1D gas is confined
longitudinally in a harmonic trap of frequency $\omega_T$, in
which case the maximum energy transfer to the BEC that can occur
is $\hbar\omega_{max}=N\hbar\omega_T$. If we arrange that the
Raman transition is detuned such that
$\hbar\omega_{max}\ll\hbar\delta$ with $\delta=c|{\bf k}_{1}-{\bf
k}_{2}|-\Delta$, then for all relevant energy transfers $k\approx
k_0=\sqrt{(2m\delta/\hbar)}$, and the implicit dependence of
$\langle\cdots\rangle_{B}^{'}$ in Eq. (\ref{FT}) on $\omega$ may
be neglected. Then the integral over $\omega$ yields a delta
function of time, and one obtains an expression for the singly
differential angular cross section in terms of the {\it static}
longitudinal momentum distribution of the target:
\begin{equation}
\frac{d\sigma}{d\Omega}\approx\beta\langle
\hat{c}^\dagger(k_0\sin(\theta),0)\hat{c}(k_0\sin(\theta),0)\rangle
_{B} , \label{cross}
\end{equation}
where $\beta$ is a constant.  Measurement of the differential
cross section as a function of angle $\theta$ out of the
transverse plane will then yield the structure of the longitudinal
momentum distribution of the target atoms. For a longitudinal trap
of frequency $\omega_T$ the characteristic longitudinal momentum
is $k_{osc}=2\pi/x_{osc}$, with $x_{osc}=\sqrt{\hbar/m\omega_T}$
the ground state width.  If $k_0>>k_{osc}$ then we may approximate
$k_0\sin(\theta)\approx k_0\theta$ for all angles of interest, in
which case the differential angular cross section will directly
mirror the momentum distribution.

{\it Evaluation of T-matrix}: Although the expressions (6) and (7)
do not require explicit evaluation of the T-matrix, it is helpful
to examine its leading term to verify that it can be absorbed into
constant prefactors. The Raman process is of second order in the
interaction between the atoms and the laser field, so we
approximate $\hat{T}$ by the second term in its Born series:
$\hat{T}\approx\hat{H}^{'}\hat{G}_{0}\hat{H}^{'}$. Here
$\hat{H}^{'}$ represents the interaction of the atomic electrons
and nuclei with the quantized electromagnetic field and
$\hat{G}_{0}=(E+i\epsilon-\hat{H}_{0})^{-1}$ with $E=E_{i}=E_{f}$
(remembering the energy delta function in the cross section) and
$\epsilon\rightarrow 0+$. $\hat{H}_{0}$ is the Hamiltonian of free
atoms and quantized electromagnetic field. We use the electric
field gauge with quantized electric field operator
\begin{equation}
\hat{{\bf E}}({\bf r})=i\sum_{{\bf k}\lambda}\sqrt{\frac{hck}
{{\cal V}}}\left({\bf e}_{{\bf k}\lambda}\hat{b}_{{\bf k}\lambda}
e^{i{\bf k}\cdot{\bf r}}-\mbox{h.c.}\right)
\end{equation}
where ${\bf e}_{{\bf k}\lambda}$ are the unit polarization vectors
and ${\cal V}$ is the quantization volume. It is assumed that the
Raman lasers are tuned close to resonance for a single S-P
transition from the atomic ground state to an electronically
excited state $\nu$ with electronic excitation energy
$\epsilon_{\nu 0}$, i.e., $\hbar ck_{1}\approx\hbar
ck_{2}\approx\epsilon_{\nu 0}$. Then the dominant contribution to
the T-matrix element in Eq. (4) is
\begin{eqnarray}
({\bf k}_{2}\lambda_{2},{\bf k}&|&T|{\bf k}_{1}\lambda_{1}, {\bf
k}-{\bf k}_{1}+{\bf k}_{2})\approx-(hc/{\cal V})\sqrt{k_{1}k_{2}}
\nonumber\\ & \times & ({\bf e}_{{\bf k}_{2}\lambda_{2}}\cdot{\bf
d}_{0\nu}) ({\bf e}_{{\bf k}_{1}\lambda_{1}}\cdot{\bf d}_{\nu 0})
/(\epsilon_{\nu 0}-\hbar ck_{1}) , \label{Tmat}
\end{eqnarray}
where ${\bf d}_{\nu 0}$ is the corresponding dipole matrix
element. The contributions $\hbar^{2}k^{2}/2m$ (kinetic energy of
ejected atom) and $\Delta$ (Zeeman splitting) to the energy
denominator have been dropped since they are much less than the
electronic transition energy. Assuming that the Raman laser
parameters are held constant, the expression (\ref{Tmat}) is
indeed a constant.

\begin{figure}
\epsfxsize 3 in \epsfbox{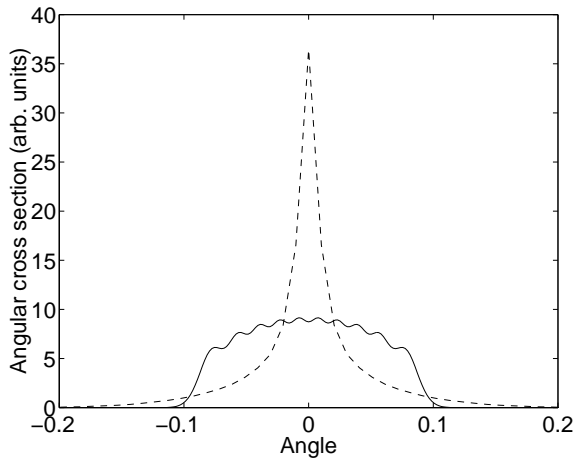} \caption{Angular
cross section versus angle $\sin(\theta)\approx\theta$ for $N=10$.
The dashed line is for the 1D gas of impenetrable bosons and the
solid line is for the corresponding system of non-interacting
fermions} \label{Fig:two}
\end{figure}

{\em Discussion}: Figure \ref{Fig:two} shows numerical results for
the angular cross section versus angle for both a 1D system of
$N=10$ hard core bosons \cite{GirWriTri01} in a harmonic trap
(dashed line) and the corresponding system of non-interacting
fermions \cite{GleWonSch00} (solid line), where we chose
$\omega_T=2\pi\times 1$ Hz, $\delta=2\pi\times10^3$ Hz, so that
$k_0=7.3k_{osc}$.

These results highlight the difference between the momentum
distributions for Fermi and Bose gases: The central peak
characteristic of a Tonks gas gets sharper as $N$ increases
\cite{Ols98,GirWriTri01}, whereas the fermionic momentum
distribution, which shows the rounded Fermi sea due to the
trapping potential with Friedel oscillations superposed
\cite{GleWonSch00}, broadens with increasing $N$. We remark again
that this difference in angular distribution arises even though
both the fermionic and bosonic atoms have the same dynamic
structure factor. This highlights the need for a measurement
technique that accesses the momentum distribution directly.  One
could measure the momentum distribution by simply releasing the
atoms from the trap and letting them expand, but this leaves open
the possibility of many-body interactions obscuring the results as
the gas expands. The momentum distribution could also in principle
be reconstructed from the single-particle reduced density matrix
$\rho_1(x,x')$ as measured in the experiment of Bloch {\it et al.}
\cite{BloHanEss00}, and it will be of interest to see how our
approaches compare.
\vspace{0.2cm}

\noindent We thank Profs. Poul Jessen and Pierre Meystre for
helpful discussions. MDG also thanks Profs. S. Stringari and L.P.
Pitaevskii for pointing out that the Tonks and Fermi gases have
the same dynamic structure factor, and for support during his stay
at the Dipartimento di Fisica, Universita di Trento. This work was
supported at the University of Arizona by Office of Naval Research
grant N00014-99-1-0806.

\end{document}